\documentclass[prb,epsf]{revtex4}

\usepackage{graphicx}
\usepackage{amsmath}
\usepackage{amsfonts}
\usepackage{mathrsfs}
\usepackage{color}
\usepackage{slashed}
\newcommand{\tn}{\textbf}

\def\<{\langle}
\def\>{\rangle}

\def\beq{\begin{equation}}
\def\eeq{\end{equation}}
\def\barray{\begin{eqnarray}}
\def\earray{\end{eqnarray}}

\newcommand{\be}{\begin{equation}}
\newcommand{\ee}{\end{equation}}

\font\numbers=cmss12
\font\upright=cmu10 scaled\magstep1
\def\stroke{\vrule height8pt width0.4pt depth-0.1pt}
\def\topfleck{\vrule height8pt width0.5pt depth-5.9pt}
\def\botfleck{\vrule height2pt width0.5pt depth0.1pt}
\def\Zmath{\vcenter{\hbox{\numbers\rlap{\rlap{Z}\kern
0.8pt\topfleck}\kern 2.2pt
                   \rlap Z\kern 6pt\botfleck\kern 1pt}}}
\def\Qmath{\vcenter{\hbox{\upright\rlap{\rlap{Q}\kern
                   3.8pt\stroke}\phantom{Q}}}}
\def\Nmath{\vcenter{\hbox{\upright\rlap{I}\kern 1.7pt N}}}
\def\Rmath{\vcenter{\hbox{\upright\rlap{I}\kern 1.5pt R}}}
\def\Pmath{\vcenter{\hbox{\upright\rlap{I}\kern 1.5pt P}}}

\begin{document}

\title{Quantum Computation of Prime Number Functions}
\author{Jos\'e I. Latorre$^1$ and Germ\'an  Sierra$^2$ \vspace{0.2cm} \\  
${}^1$ Departament d'Estructura i Constituents de la Mat\`eria, Universitat de Barcelona, 
Barcelona, Spain,   \\ 
Centre for Quantum Technologies, National University of Singapore, Singapore.\\
 ${}^2$  Instituto de F\'isica Te\'orica UAM/CSIC, Universidad Aut\'onoma de Madrid, Cantoblanco, 
 Madrid, Spain.}

\begin{abstract}
We propose a
quantum circuit that creates  a pure state corresponding to the quantum  superposition of 
all prime numbers less than  $2^n$, where   $n$ is the number of qubits of the register. 
This {\em Prime} state can be built
using Grover's  algorithm,  whose  oracle is a quantum implementation  of the classical Miller-Rabin primality test.  
The {\em Prime} state  is highly entangled, and its entanglement measures  encode 
number theoretical functions such as the distribution of twin primes or the Chebyshev bias. 
This algorithm can be further combined with the quantum Fourier transform to yield an estimate  of the prime counting function, more efficiently than any classical algorithm and with an error below the bound that allows for the verification of the Riemann hypothesis.  We also propose 
a {\em Twin Prime}  state to measure the number of twin primes and another state to test the Goldbach conjecture. 
Arithmetic properties of  prime numbers are then, in principle, amenable to experimental verifications on quantum
systems. 
\end{abstract}


\maketitle

\section{Introduction}

Prime numbers are central objects in Mathematics and Computer Science. 
They appeared  dramatically  in Quantum Computation through the
Shor's algorithm,  which converts  the hard problem of factorization into
a polynomial one using quantum interference \cite{S97,NC}. 
In  Number Theory,   prime numbers are fully  characterized by the prime
counting function $\pi(x)$, which is the number of primes less or equal to $x$. 
This is a stepwise function which jumps by one whenever $x$ is a prime. For example
$\pi(100) = 25$ means that there are 25 primes below or equal to 100, but
$\pi(101) = 26$ because 101 is a prime.  The asymptotic behavior of $\pi(x)$ 
is given by the Gauss law $\pi(x) \sim  {\rm Li}(x)$, where ${\rm Li}(x)$ is the logarithmic
integral function, which for large values of $x$ behaves as
$ x/\log x$ \cite{D}. This statement
is known as the Prime Number Theorem (PNT). Moreover, the fluctuations
of $\pi(x)$ around ${\rm Li}(x)$,  will be  of order $\sqrt{x} \log x$,
if and only if  the Riemann hypothesis holds true \cite{D}.  Other interesting number theoretical functions
are $\pi_k(x)$ which gives the number of primes $p \leq x$, such that
$p+ k$ is also a prime. In particular, the function $\pi_2(x)$ counts the number
of twin primes. According to a famous conjecture due to Hardy and Littlewood, 
 $\pi_k(x) \sim  2 C_k \,  x/(\log x)^2$  for $x \gg 1$ \cite{HL}, where $C_k$ is a $k$-dependent constant. 

The aim of  this paper is to  show that  the number theoretical functions $\pi(x), \pi_k(x)$
and others,  can be computed in an efficient way using quantum entanglement
as the main  computational  resource.  
In our approach, prime numbers are represented by
quantum objects which are treated  as a whole with  the computational tools 
provided by  spins, photons, ions,  or  other quantum devices. The  results we obtain  suggest that 
difficult number theoretical problems could be addressed
experimentally,  once large scale quantum computation becomes available.

\section{The Prime state}

Our  starting point is the  {\sl Prime}  state  made of $n$-qubits that corresponds to the
quantum superposition of all prime numbers less than $2^n$ (we take  $n >1$ so that  $2^n$ is not a prime),
\begin{equation}
  | \Pmath_n \rangle \equiv  \frac{1}{\sqrt{\pi(2^n)}}\sum_{p\in {\rm primes} < 2^n}| p\rangle \ ,
\label{primestate}
\end{equation}
where each prime number can be expressed in binary form $p=p_0 2^0+p_1 2^1+\ldots +p_{n-1} 2^{n-1}$, and is then translated into a quantum register on the computational basis $|p\rangle=| p_{n-1},\ldots,p_{1}, p_0\rangle$. Note that all the states in the sum are orthogonal and that the normalization of the state is  related to the
squared root of the number of primes less than $2^n$, namely $\pi(2^n)$.

As an example consider the case of $n=3$. Then
\begin{eqnarray}
  |\Pmath_3  \rangle &=& \frac{1}{\sqrt{4}} \left( 
   |2\rangle + |3\rangle + |5\rangle + |7\rangle \right) \label{example}  \\
   &=& \frac{1}{2} \left(
   |\uparrow  \downarrow \uparrow \rangle + |\uparrow \downarrow  \downarrow \rangle + 
   |\downarrow \uparrow  \downarrow\rangle + |\downarrow \downarrow \downarrow \rangle  \right). 
\nonumber 
\end{eqnarray}
where the qubits $|0\rangle$ and $|1\rangle$ are described by the spin polarized states  $\uparrow$ and $\downarrow$   
of a spin $1/2$ particle. Other physical realizations of qubits are of course equivalent.

Several questions arise regarding the {\em Prime}  states: 
{\sl i}) how to prepare  them, 
{\sl ii}) how to compute the functions $\pi(2^n), \pi_k(2^n)$, etc,
{\sl iii}) what are their entanglement properties, and 
{\sl iv}) are there Hamiltonians whose  ground states are $|\Pmath_n  \rangle$. 
These questions will be answered below
combining  standard methods in  Quantum Computation and Number Theory. 

The answer to  question {\sl iv}) can be readily given. 
It is just sufficient to take as Hamiltonian any primality test 
algorithm that, acting on an integer $x$, returns a zero for prime numbers and any positive eigenvalue 
$\lambda_x$ for composite numbers, that is
\begin{eqnarray}
  \nonumber
  &H_{\rm primality} |x\rangle = 0 \qquad&{\rm if\ } x\in \Pmath, 
  \\
  &H_{\rm primality} |x\rangle = \lambda_x |x\rangle \quad &{\rm if\ } x\notin \Pmath, 
\label{primalitymailtonian}
\end{eqnarray}
where $\Pmath = \Pmath_\infty$ denotes the set of all  prime numbers. 

A more relevant approach consists in turning a classical primality test algorithm into a  quantum circuit $U_{\rm primality}$ that is capable of discriminating  prime from composite numbers
\be
  U_{\rm primality} \sum_{x=0}^{2^n-1}| x\rangle |0> =|\Pmath_n \rangle   |0> + A \sum_{c \in {\rm composite}}
    |c> |\lambda_c\rangle \ ,
\label{primalitycircuit}
\ee
where the ancilla $|\lambda_c\rangle \not= |0\rangle$, $A$ is a normalization constant and 
the explicit construction of an example of $U_{\rm primality}$ will be presented later on.
It is then possible to create the {\em Prime}  state by performing a measurement of the ancilla.
The probability to project onto the {\em Prime}  state is
given by the probability of measuring 0 on the ancilla register,
\be
  {\rm Prob}(| \Pmath_n  \rangle)=\frac{\pi(2^n)}{2^n}\sim \frac{1}{n \log 2}, 
\label{probprime}
\ee
where we have used the PNT, which shows the  efficiency of the algorithm, since 
the probability to obtain the Prime state is only polinomially suppressed.

As a result, we may argue that this circuit brings the possibility of
measuring $\pi(2^n)$. It is enough to repeat the preparation and keep the 
statistics  of the output for the ancilla measurement.
Even though  the circuit is efficient, it shares the same complexity as a
classical computer trying to assess the value of $\pi(2^n)$. However, conceptually the 
two approaches are quite  different. On a classical computer  every time
we create a number, and test for primality,  we simply get one prime number or none. 
Instead, the quantum circuit creates the superposition of all primes.
This allows for the {\em Prime}  state to be further used 
to explore the distribution of prime numbers.
We shall show later that there is a more efficient method to create and
analyze the {\sl Prime state}, using a combination of a quantum oracle for primality
and the Quantum Fourier Transform.

\section{Twin primes and Goldbach conjecture}

The construction of the  {\em Prime}  state can be generalized in a straightforward  way
to states that  encode important concepts and  problems in Number Theory. 
Let us start with a  very simple circuit that checks for twin primes.
Consider creating the prime state, and then adding 2 to each prime
\be
  U_{+2}| \Pmath_n\rangle = \sum_{p\in {\rm primes}< 2^n} |p+2\rangle
\label{twin1}
\ee
We then act again with the basic primality circuit
\be
  U_{\rm primality} \sum_{p:{\rm primes} < 2^n} |p+2\rangle  |0 \rangle =
  A \sum_{q\in{\rm primes}<2^n} |q \rangle  |0\rangle + B \sum_{c\in{\rm composite}<2^n} 
  |c\rangle |\lambda_c\rangle
\label{twin2}
\ee
When measuring the ancilla, the probability of finding a prime which is
twin of a previous prime is
\be
  {\rm Prob}\left((p,p+2)\in {\rm primes}\right)= \frac{|A|^2}{\pi(2^n)}
\label{twin3}
\ee
On the other hand, this probability is given by the ratio 

\be
  {\rm Prob}\left((p,p+2)\in {\rm primes}\right)= \frac{\pi_2(2^n)}{\pi(2^n)}
\label{twin3b}
\ee
where $\pi_2(x)$ is the counting function for twin primes below or equal to $x$.
Using the Hardy-Littlewood conjecture the ratio (\ref{twin3b})  has $O(1/n)$. 
 Given that the production of
the Prime state is itself supressed by a factor $1/n$, the global probability 
of measuring twin primes experimentally is expected to be $1/ n^2$. This matches the
same difficulty as computing the density of twin primes on a classical computer.

It is also possible to create a circuit that tests the Golbach conjecture,
which states that every even integer greater that 2 can be expressed 
as the sum of two primes. Excluding the case $4= 2+2$, one
can formulate this conjecture saying that every even integer greater than
4 can be written as the sum of two odd primes. The first case being given
by $6=3+3$.  To formulate the Goldbach conjecture in Quantum Mechanics we shall
define the state associated to  odd prime numbers 

\begin{equation}
  | \Pmath_{{\rm odd},n}  \rangle = \frac{1}{\sqrt{\pi(2^n) -1}}\sum_{2 < p  < 2^n}| p\rangle \ ,
\label{goldbach0}
\end{equation}
where the summation is restricted to odd prime numbers less than  $2^n$.
Consider now the creation of a product state of two odd Prime states, and apply 
a sum operation 
\be
|{\rm Goldbach}_n \rangle  \equiv 
  U_{+} \left( | \Pmath_{{\rm odd},n}  \rangle  | \Pmath_{{\rm odd},n}  \rangle \right) = \frac{1}{\pi(2^n) -1}
  \sum_{(p,q)\in {\rm odd \; primes} < 2^n} |p\rangle |p+q\rangle \ .
\label{goldbach1}
\ee
This circuit  puts on the second register the addition of two odd  primes.
The state on the RHS uses a register with $2 n +1$ quits.
The reason being  that the sum of two numbers between $0$ and $2^n-1$
runs up to $2^{n+1}-2$, so $n+1$ digits are required to store the result,
which added to the $n$ qubits for the first register gives $2n+1$. 

The sum $p+q$ is an  even number greater or equal to 6. The Goldbach
conjecture asserts that all the even numbers will appear in the second register
of (\ref{goldbach1}) for sufficiently large values of $n$. 
Again, this strategy does not bring any improvement over a classical 
strategy but is conceptually different since the second register
contains the superposition of all even numbers.

\section{Entanglement of the Prime state}

The {\em Prime}  state must carry a large amount of quantum entanglement.
Otherwise, it would be possible to simulate it on a classical 
computer with polynomial resources.  
A good figure of merit to quantify the entanglement present in the {\em Prime}  state 
is the von Neumann entropy for the reduced density matrix of 
a subsystem. To be concrete, we first divide the system in the first $l$ qubits
and the rest $n-l$ qubits. Then the reduced density matrix
\be
  \rho(l)={\rm Tr}_{n-l} |\Pmath_n \rangle\langle \Pmath_n| , 
\label{rho}
\ee
is computed. Finally we calculate the entanglement entropy
\be
  S\left(\rho(l)\right)=-{\rm Tr}_l \rho(l) \log \rho(l) .
\label{entropy}
\ee
There are two relevant properties of the von Neumann entropy
of the {\em Prime}  state. First, we fix the size of the register  $n$
and we observe that  the entropy grows
approximately as $\log \, l$.  
Second, we consider the even bi-partition of the system $l=n/2$, with $n$ even, and explore
how the entropy $S(\rho(n/2))$ varies with  $n$. The entropy can be seen to clearly scale {\em almost}  in the  maximal way, that is linearly in $n$. Both results have been obtained from exact numerical simulations up to $n=22$.

An interesting question is how single  qubits are entangled with the 
rest of the qubits in the {\em Prime}  state. This is described by the reduced density matrices 
\be
\rho^{(i)}  = {\rm Tr}_{n/ i}  \, |\Pmath_n \rangle\langle \Pmath_n|, \quad i=0, 1 \dots, n-1, 
\label{na5}
\ee
where the trace excludes the  $i^{\rm th}$ qubit. For the last qubit,  $i=0$, one finds  
\be
 \rho^{(0)}_{ 0,  0}  = \frac{1}{\pi(N)},   \rho^{(0)}_{ 1,  1}  = \frac{\pi(N) -1  }{\pi(N)},    \rho^{(0)}_{ 0,  1}  = \frac{1  }{\pi(N)},
\label{na10}
\ee
where $N= 2^n$. For a large number of qubits $n$, 
the PNT implies that   the  entanglement  entropy of the last  qubit  decreases exponentially with $n$
\be
S_0 = - {\rm Tr} \, \rho^{(0)}  \, \log \, \rho^{(0)}  \sim  2^{-n} (n \log 2)^2. 
\label{na13}
\ee
The reason being  that all the primes but 2 are odd, so the last  qubit  is mostly in the state
$p_0=1$.
A more interesting  result is obtained for the next to last qubit, $i=1$,  whose density matrix is 
\be
 \rho^{(1)}_{ 0,  0}   = \frac{\pi_{4,1}(N)}{\pi(N)},   \rho^{(1)}_{ 1,  1}  = \frac{1+ \pi_{4,3}(N)   }{\pi(N)},  \rho^{(1)}_{ 0,  1}  = 
 \frac{\pi_2^{(1)}(N)  }{\pi(N)},
\label{na16}
\ee
where $\pi_{a,b}(x)$ is the number of primes less or equal to $x$ that appear in the arithmetic
progression $a m + b$, with $a$ and $b$ coprime numbers, 
and $\pi_2^{(1)}(x)$ is the number of prime pairs $(p, p+2)$ less or equal to $x$ with  $p= 1 \,  {\rm mod } \, 4$.
There are also prime pairs with $p= 3 \,  {\rm mod } \, 4$, in number $\pi_2^{(3)}(x)$, but they do not contribute
to $\rho^{(1)}_{ 0,  1}$. The sum $\pi_2^{(1)}(x) + \pi_2^{(3)}(x)$ is equal to the 
 twin primes counting function $\pi_2(x)$. 
Dirichlet proved that the number of primes in  these arithmetic progressions is infinite \cite{D}.
Furthermore,  the  fraction of these primes relative to the total number of primes satisfies
a version of the PNT, 
\be
\lim_{ x \rightarrow \infty} \frac{ \pi_{a,b}(x)}{ {\rm Li}(x)} = \frac{1}{ \phi(a)}
\label{na15}
\ee
where $\phi(a)$  is the Euler totient function,  which is the number of positive
integers $x <  a$ which are relative prime to $a$. 
Using this result and the fact that $\phi(4) = 2$, one finds that the entanglement
entropy of the  qubit $i=1$  behaves asymptotically as 
\be
S_1 = - {\rm Tr} \, \rho^{(1)}  \, \log \, \rho^{(1)}  \sim \log 2, \qquad n \gg 1.
\label{na16}
\ee
So this qubit is maximally entangled with the rest. The same property holds for 
the remaining qubits. The reduced density matrices  $\rho^{(i)}$ also provide  the expectation values of
 local operators in  the {\em Prime}  state. In particular for  the Pauli  matrices $\sigma^a_i$ one has 
 \be
 \langle \sigma^a_i \rangle = {\rm Tr} \, (  \rho^{(i)}  \sigma^a_i ), \quad a=x,y,z, \quad i=0, \dots, n-1. 
 \label{na26}
 \ee 
For  example the magnetization of the qubit $i=1$  reads
 \be
  \langle \sigma^z_1 \rangle  =  
 \frac{ \pi_{4,1}(N)  -  \pi_{4,3}(N) -1  }{\pi(N)}. 
\label{na28}
\ee
 The numerator is essentially the Chebyshev bias \cite{che} 
 \be
 \Delta(x) = \pi_{4,3}(x)  -  \pi_{4,1}(x),
  \label{na29}
 \ee
 which counts the unbalance of the remainder upon dividing a prime by 4. For low values
 of $x$, the remainder 3 appears more often than the remainder 1, but Hardy and Littlewood
 showed that the relative size of  $\pi_{4,3}(x)$ and $\pi_{4,1}(x)$ vary  infinitely often so that $\Delta(x)$
 can be either positive or negative \cite{che}. 
 This  result,  known as the prime quadratic effect, 
 could be observed experimentally by measuring
 $\langle \sigma^z_1 \rangle$.   Similarly, the twin prime functions $\pi_2^{(1,3)}(N)$
 are the expectation values of one and two  sites spin flips operators, 
  \be
  \langle \sigma^x_1 \rangle  =  
 \frac{ 2 \,  \pi_{2}^{(1)}(N)  }{\pi(N)}, \quad   \langle \sigma^x_1 \sigma^x_2+  \sigma^y_1 \sigma^y_2 \rangle  =  
 \frac{ 4 \,  \pi_{2}^{(3)}(N)  }{\pi(N)}. 
\label{flip}
\ee
In analogy with eq.(\ref{na29})
we  can define the twin prime bias $\Delta_2(x) =  \pi_{2}^{(3)}(x) -  \pi_{2}^{(1)}(x)$,
which seems also to oscillate. 
\bigskip

\section{Primality quantum oracle}

A different way to prepare the {\em Prime}  state corresponds to use a primality 
module as an oracle in Grover's  algorithm \cite{G}. We are searching for 
$M= \pi(2^n)$ items (the primes below $2^n$)  within a set of $N = 2^n$ objects
(the integers between 0 and $2^n-1$).  Using the Grover's algorithm, on a  quantum computer,  
this search  can be performed
in $O(\sqrt{N/M})$ steps with a high probability, which represents
a significant computational gain \cite{NC}.

\begin{figure}
\centering
\includegraphics[width=0.4\textwidth]{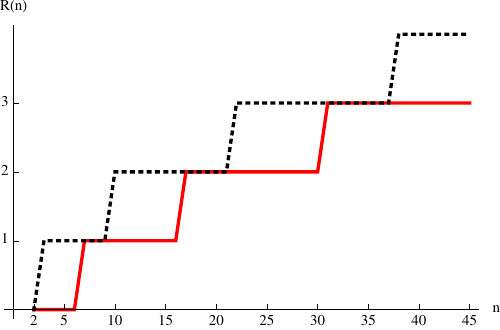}
\hspace{0.5cm}
\includegraphics[width=0.4\textwidth]{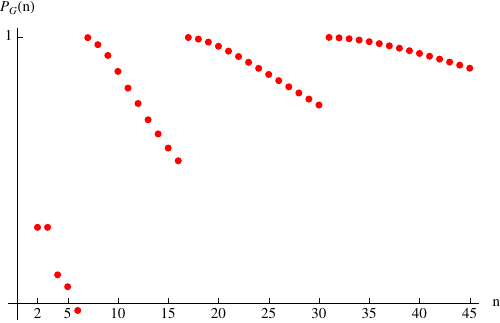}
\caption{Left: Number of Grover's steps needed to create the {\em Prime} state $|\Pmath_n \rangle$ in the range $n=2, \dots, 45$
($R(n)$: continuous line  and $R_{\rm max}(n)$: dashed line). 
Right: Accuracy of the state measured by  $P_G(n)$. } 
\label{R-Grover} 
\end{figure}

As  oracle for the Grover's algorithm we use the unitary transformation 
\be
  U_f |x\rangle = (-1)^{f(x)} \;  |x\rangle
\label{g7}
\ee
where $f(x)=1$ if $x\in \Pmath_n$ and $f(x)=0$ if  $x \notin \Pmath_n$. 
 One next  introduces the  unitary  $U_\psi = 2 | \psi \rangle \langle \psi| - {\bf 1}$, 
where $|\psi \rangle = N^{-1/2}   \sum_{x=0}^{N-1} |x \rangle$ is the state obtained
applying $n$ Hadamard transforms to the initial state $|0 \rangle^{ \otimes n}$.
Grover's transformation, defined as  $G = U_\psi  U_f$,  is  applied iteratively to the state $|\psi \rangle$
until it gets closed to  the target state $|\Pmath_n\rangle$. 
The optimal value of iterations, $R(n)$,  is estimated by 
\be
R(n) = 
\left[   \frac{   \arccos \left( 2^{- n/2} \sqrt{ \pi(2^n)} \right)  }{ 2  \arcsin  \left( 2^{- n/2} \sqrt{ \pi(2^n)} \right)   }  \right] 
\label{g14}
\ee
where $[ x ]$ denotes the integer part of $x$. If $M \leq N/2$, as it occurs in our problem, there is  an upper bound
%
\be
R(n) \leq \left[ \frac{ \pi}{4} \sqrt{ \frac{N}{M} } \right]  \leq R_{\rm max}(n)  \equiv   \left[ \frac{ \pi }{4} \sqrt{ n \log 2}   \right]
\label{g15}
\ee
which follows from  the PNT for $n \gg 1$.  Hence the Grover's algorithm requires $O(\sqrt{n})$ calls
to the oracle, which represents a computational gain compared to a classical algorithm (see Fig. \ref{R-Grover}). 
Note  that one needs about 3 Grover's iterations to construct an  approximation to the  {\em Prime}  state
up to $2^{45}  \sim 3.5 \times 10^{13}$!! To assess the goodness of the approximation
we compute  the overlap between 
the {\em Prime}  state with the Grover state after $R(n)$ iterations
\be
P_G(n) = | \langle  \Pmath_n | G^{R(n)} \, |\psi \rangle |^2  = \sin^2 \left[  \frac{ (2  R(n) + 1)  \theta(n)}{2}  \right] 
\label{g18}
\ee
where $\theta(n)$ is the Grover's angle 
\be
\theta =  \theta(n) = 2 \arcsin  \sqrt{M/N}  = 2 \arcsin \left( 2^{- n/2} \sqrt{ \pi(2^n)} \right). 
\label{g6}
\ee
The overlap (\ref{g18}), shown in Fig.\ref{R-Grover}, has some jumps with $n$  but it approaches 1 rather fast as  $n$  increases. 
 
 %
%

The above Grover construction relies on the fact that some classical primality
tests  can be turned into a quantum oracle. This is the case of the 
Miller-Rabin primality test which we will write down below as  a quantum circuit. 
The remarkable AKS primality test (\cite{AKS}, which is unconditional, deterministic and efficient) 
could also be turned into an oracle. Nevertheless, the Miller-Rabin test
has a simpler structure which makes easier its conversion into a 
quantum primality oracle.

Let us first summarize the Miller-Rabin primality test \cite{MR}. 
The goal is to declare a number $x$  either prime or composite. First, it is necessary to find the integers 
$s$ and $d$ (odd) such that an odd number $x$ is decomposed as $x-1=2^s d$. We then choose a number $a$, 
in the range $1 \leq a < x$,  that is called witness and check 
\begin{eqnarray}
 &&a^d\not \equiv  1\ \pmod x \quad   \label{MR} \\
 &&a^{2^r d} \not \equiv -1 \pmod x\qquad 0\le r\le s-1. \nonumber 
\end{eqnarray}
If all these tests are verified, $x$ is composite with certainty. 
However if the test fails, $x$ can be either prime or composite.  
In the latter case  the number $a$ is called a strong liar to $x$.
In order to circumvent strong liars, it is necessary to rerun the test with  different witnesses. 
As more witnesses are tested, the probability to be deceived by strong liars vanishes. 
Assuming the Generalized Riemann Hypothesis (GRH), the Miller-Rabin test is deterministic using less than $\log^2 x$ witnesses. 
For instance, all numbers below $x<3\ 10^{14}$ can be correctly classified as prime or composite using as witnesses $a=2,3,5,7,11,13,17$.
We can also implement the probabilistic version of the Miller-Rabin test which does not assume the GRH
and that using  $k$ witnesses declares a composite to be prime with an error less than $2^{- 2k}$ \cite{MR}. Hence  choosing
$k$ to be equal to $n$ the error will be negligable for our purposes. 

\begin{figure}
\centering
\includegraphics[width=0.8\textwidth]{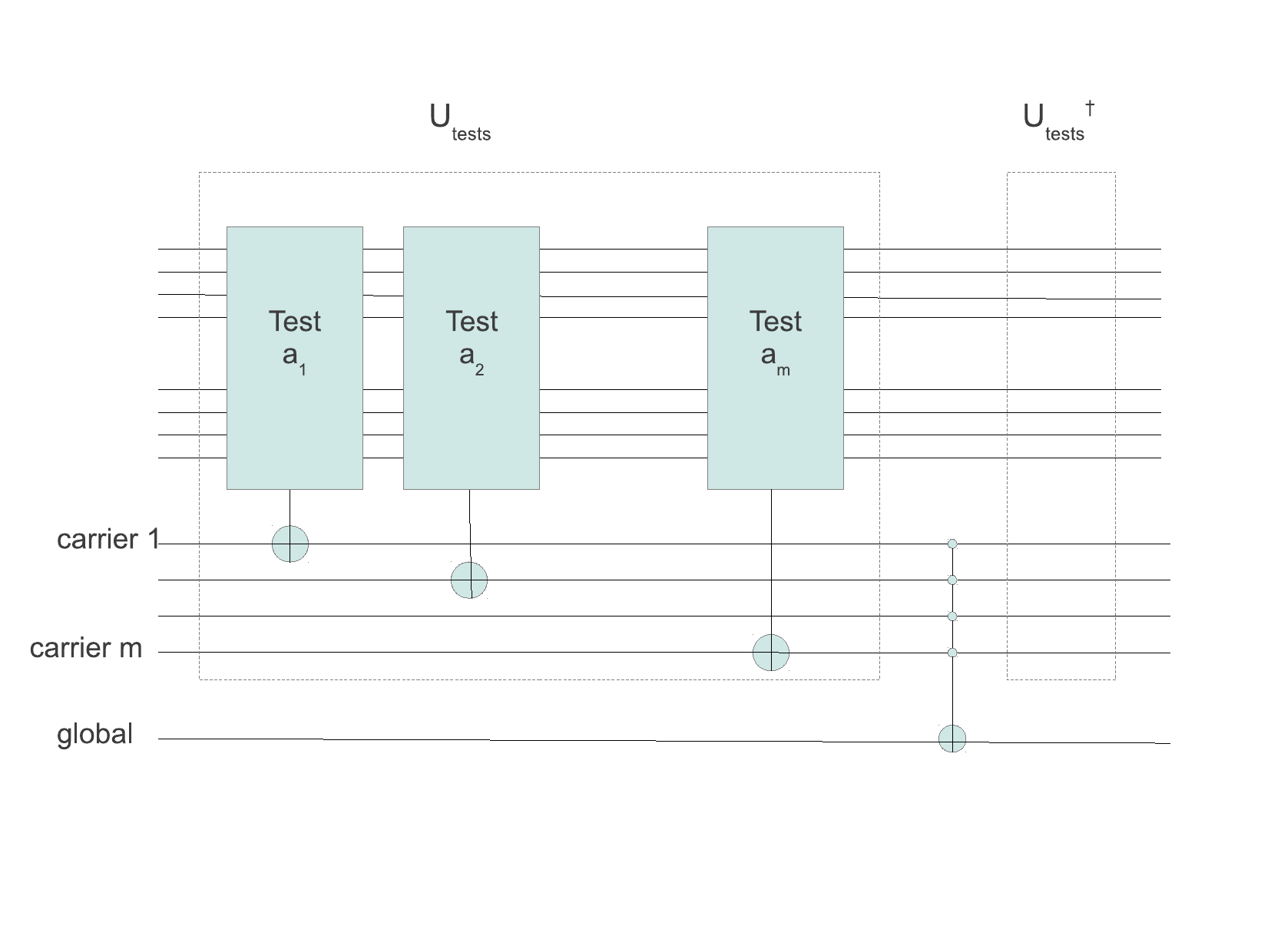}
\caption{Structure of the Quantum Primality Oracle based on the Miller-Rabin primality test. A series of unitary modules implement a quantum version of the modular exponentiation tests required by the classical test. The total number of tests $m$ is smaller than $n^2$.}
\label{globaloracle} 
\end{figure}

\begin{figure}
\centering
\includegraphics[width=0.7\textwidth,angle = 90]{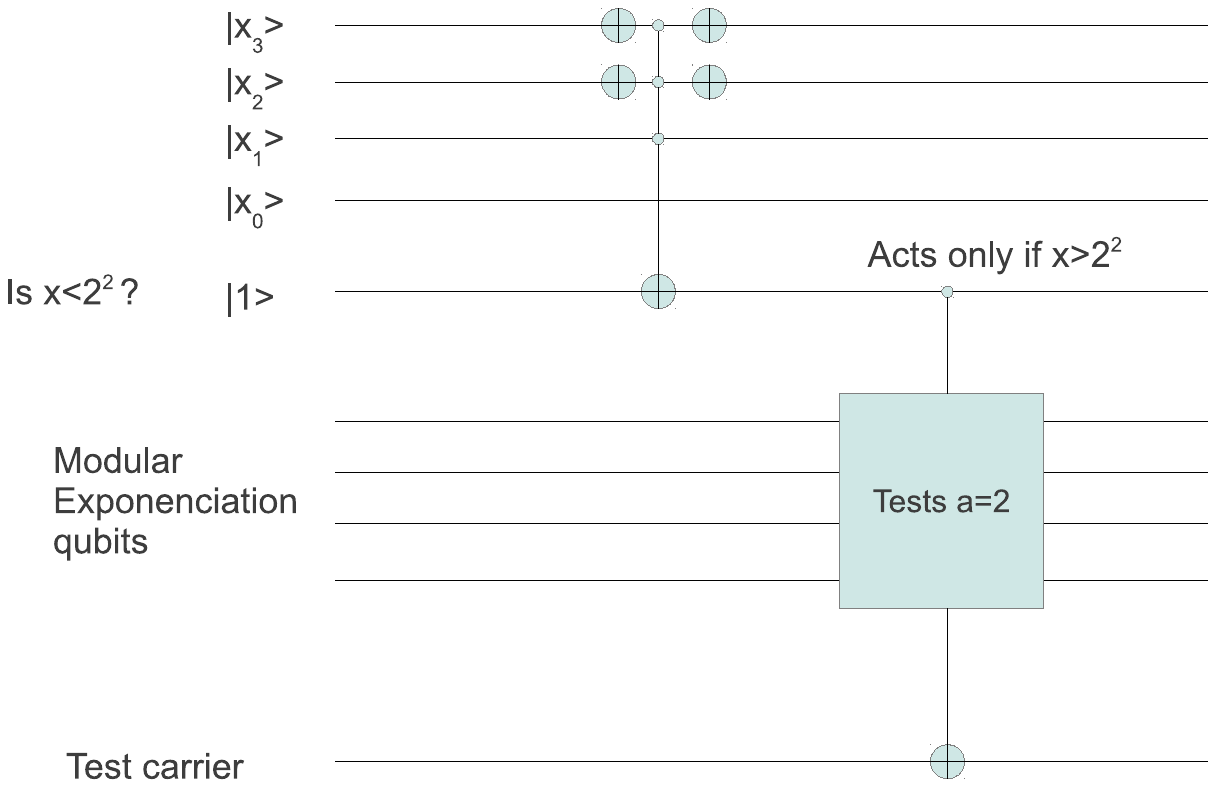}
\caption{Detail of the condition that guarantees that only those witnesses less than $x$ are tested. 
} 
\label{oracle1} 
\end{figure}

The quantum primality oracle based on the Miller-Rabin test follows closely the steps of
the classical test. Basically, a series of unitary modules implement a quantum version of each of 
the classical modular exponentiation tests that form the Miller-Rabin test, as shown in Fig. \ref{globaloracle}.
In order to simplify
the algorithm, we shall consider the construction of the odd superposition in the {\em Prime} state, that is
 we leave out the element $| 2 \rangle$. This is a trivial element that could be restored with a simple 
initial controlled gate. We thus start by preparing the superposition of all odd numbers less than $2^n$, using $n-1$ Hadamard operations on the first $n-1$ qubits, while leaving the last
qubit set to $|x_0\rangle= |1\rangle$, and adding a set of target ancillae that will be used to implement the modular exponentiation tests
\be
  |\psi_0\rangle =\frac{1}{2^{(n-1)/2}}\sum_{x_{n-1},\ldots,x_1=0,1}|x_{n-1},\ldots,x_1,1\rangle|0\rangle.
\label{inicialstate}
\ee
For each value of $|x\rangle$ we need to find two states $|d\rangle$ and  $|s\rangle$ such that  $|x-1\rangle= |d\rangle |s\rangle$. This can be done using the fact that $|s\rangle$  is related to the number of trailing zeros in the register when subtracting 1, while $|d \rangle$ is related to the initial set of the qubits. Let us illustrate this fact in the case where the register reads $|25\rangle=|1,1,0,0,1\rangle$, where we have $|d\rangle=|3\rangle=|11\rangle$ (from the initial $|1,1\rangle$ piece of the register) and $s=3$ (that is $|000\rangle$ from the trailing $|0,0,1\rangle$ minus 1, see Fig. \ref{oracle1}).
This example shows that a series of gates controlled by  several qubits is enough
to perform the modular exponentiation as a unitary operation
\be
  U_{a, r}\sum_x |x\rangle |0\rangle = \sum_x |x\rangle |a^{2^r   d} ({\rm mod}  \,  x) \rangle, \quad   0\le r\le s-1. 
\label{modularexponenciation}
\ee

Nevertheless there is a subtle detail to be considered.
The Miller-Rabin test requires the witness $a$ to be smaller than $x$. 
Fortunately, this  condition also
guarantees that the above operation is unitary \cite{E}.  
Therefore the action
of each unitary modular exponentiation needs to check that $x$ is large enough for
each witness. This again is simply taken care of by a controlled gate to the most 
relevant qubits in $x$ (see Fig. \ref{oracle1}). For instance, a gate control to the most relevant qubit in $x$,
that is $|x_n\rangle$ will act when the qubit is $|1\rangle$, that is when $x>2^{n-1}$ and
all witnesses less than $2^{n-1}$ can be used. Let us here recall that the values of the
witnesses in the Miller-Rabin  algorithm are far smaller than the values of $x$ they can
test.

\begin{figure}
\centering
\includegraphics[width=0.8\textwidth, angle =90]{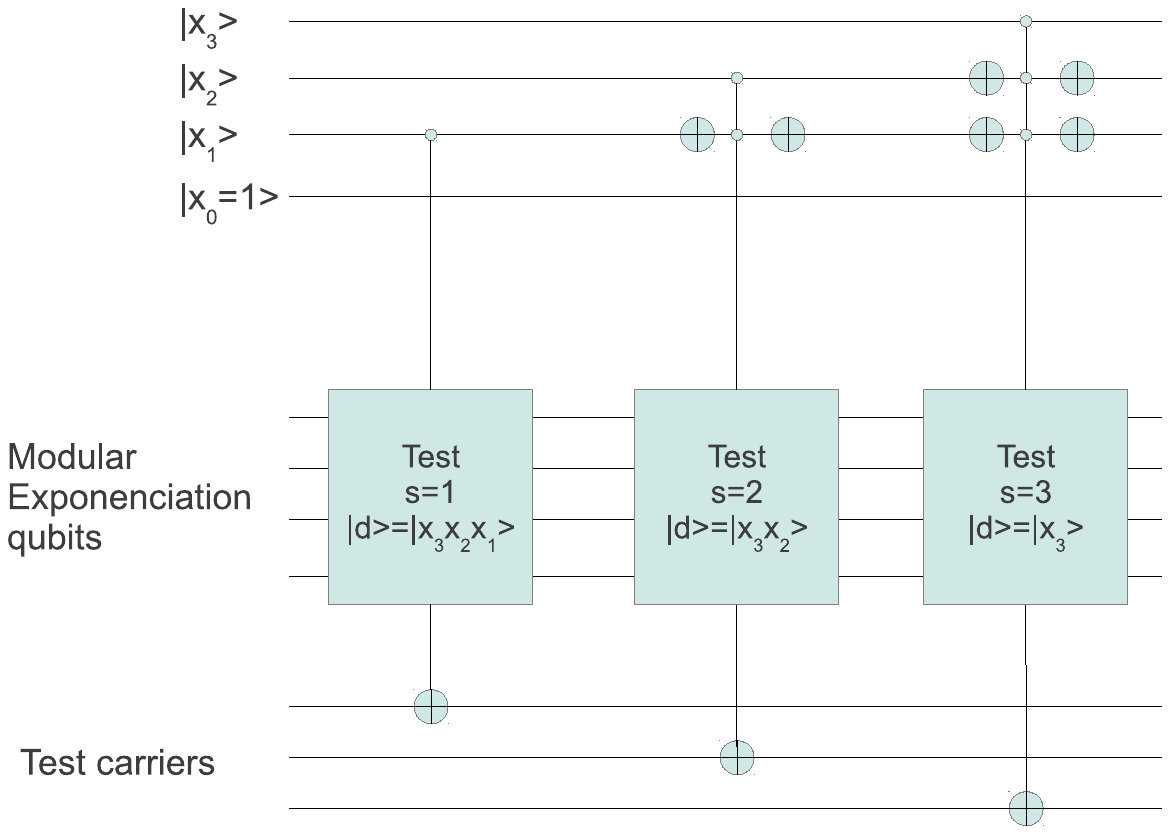}
\caption{Implementation of $x-1=d 2^{s}$ on a quantum circuit. The value of $s$ is inferred by the trailing zeros in the register, while $d$ is read from the leading qubits. Each test result is retained in its corresponding carrier. All carriers will be later collect into a single global one. } 
\label{oracle2} 
\end{figure}

The next step in the algorithm is to collect the result of the tests. The guiding principle is
to assume $x$ is composite till proven prime. A set of ancillary carriers will be 
initialized in a state corresponding to {\em composite} 
unless they are changed by the result of a test, which will correspond to {\em prime}. Let us recall that
in order for an integer $x$ to be declared a probable prime we just
 need to find that $a^d=1$ or  $a^{2^r d}=-1$ for some $r \in [0, s-1)]$ (see Eq.(\ref{MR})).
This can be achieved quantum mechanically by adding an ancillary carrier for each test based on the witness $a$ and the value $r$, that we call test$_{a,r}$ such that it is initially set to $|{\rm test}_{a,r}\rangle=|1\rangle$, and  then changed to $|0\rangle$ if the  test$_{a,r}$ fails detecting  a probable prime, 
\be
  U_{{\rm test}_{a,r}} |a^{2^r d } \,  ({\rm mod} \,   x) \rangle |0\rangle= |a^{2^r d} \,  ({\rm mod} \,   x) \rangle |{\rm test}_{a,r}\rangle \ .
\ee
After the action of all tests, all ancillae carriers will be $|1\rangle$ only for composite numbers, and will have at least one $|0\rangle$ for prime numbers. This suggests to include
a single extra global ancilla that summarizes all tests, 
initialized to $|1\rangle$. We then perform a 3-body gate $U_{\rm global}$ controlled to all test ancillae displayed in Fig. \ref{oracle2}
\be
  U_{\rm global} \prod_ {a,r} | {\rm test}_{a,r}\rangle |1\rangle =
 \prod_{a,r} | {\rm test}_{a,r} \rangle | {\rm global}\rangle
\ee
 If $x$ is prime, the {\em global}  ancilla will flip to $|0\rangle$, and  if $x$ is composite it will remain in the state $|1\rangle$. This is precisely what we need to implement the Grover condition in the usual way using a single state. 

To finish the algorithm, after the Grover sign flip on primes is achieved, we need to invert all the unitary operations so as to reset all ancillae
to their initial product state.

The computational complexity of the Miller-Rabin quantum oracle is only polynomial.
We can bound the number of basic operations in the following way. There are at most $n^2$ 
witnesses to be tried. Each witness needs at most $n$ exponential tests. Each test is of order
$n^3$ operations. Altogether, the oracle complexity scales as $n^6$. This counting assumes that
some test carriers and control operations to guarantee that $a<x$ are done using single
Toffoli-like gates. Note that, 
as a matter of fact,
the number of witnesses needed in practice is  lower than the $n^2$ bound proven using the Generalized Riemann hypothesis. 
Therefore, the algorithm will work in a faster way in practice.

\section{Quantum counting of prime numbers}

The power of the Grover algorithm becomes manifest  when it is combined with the
efficient quantum Fourier transform. The Quantum Counting algorithm \cite{BHT} is based on
the idea that the Grover module is followed by an appropriate controlled phase gate in such
a way that, after completion of the series of calls to the oracle, a quantum Fourier transform is
performed to read the number of solutions to the oracle. 

In our case, the Quantum Counting algorithm
that makes use  of our Grover primality oracle
allows to compute the number of
solutions $\pi(x)=\pi(2^n)=M$ with a bounded error. To be precise, it will produce an estimate $\tilde M$ to the actual number of solutions $M$ to the oracle such that
\be
  \left|\tilde M- M\right|< \frac{2 \pi}{c}  M^{1/2}+ \frac{\pi^2}{c^2}, 
\label{quantumestimate}
\ee
where  $c$ is a constant, 
using only $c \,  N^{1/2} =c \,  x^{1/2}$  calls to the oracle, that is time steps. 
Given that $\pi(x)\sim x/\log x$, our quantum algorithm can verify the
prime counting function with an accuracy
\be
   \left|\tilde\pi(x)-\pi(x)\right|< \frac{2 \pi}{c}  \frac{x^{1/2}}{\log^{1/2} x}
\label{quantumestimate2}
\ee
and  $O(n=\log(x))$ space allocation.

These results  provide an exponential
gain with respect to known  classical algorithms, when considering the need for both time and memory resources.
The classical computation of $\pi(x)$ of use was proposed by 
Lagarias, Miller, and  Odlyzko \cite{LMO},  who  refined the Meissel-Lehmer method. 
The number of bit operations is of order $x^{2/3}$ and the storage needed is 
of order $x^{1/3}$, where both scalings have log corrections. Lagarias and Odlyzko
have also proposed two  analytic $\pi(x)$-algorithms based on the Riemann zeta
function, whose order in  time and space  are $x^{ 3/5 + \epsilon} \; (\epsilon >0)$ and  $x^{\epsilon}$
in one case,  and $x^{ 1/2 + \epsilon}$ and  $x^{1/4+\epsilon}$ in the other case \cite{LO}. 
The latter algorithms  has been  implemented numerically to compute $\pi(10^{24})$ unconditionally \cite{P}.  Classically, it is possible to 
find other algorithms that trade space with time, yet the product of time and memory is always
bigger than order $x^{1/2}$.  The estimation of $\pi(x)$,  given by eq.(\ref{quantumestimate2}), 
is smaller than the error predicted under  the Riemann hypothesis (RH), i.e.
$|\pi(x) - {\rm Li}(x) | < O( \sqrt{x} \log x)$, thus the RH  could be falsified 
experimentally on a quantum computer with  numbers far beyond the reach on any classical computer.
However,  the proof of the RH cannot be achieved using  this method since
that would require testing systems of arbitrary size.  

\section{Conclusion}

We have shown that the quantum superposition of states that codify prime numbers 
in the computational basis, the {\sl Prime} state, can be created efficiently using a quantum circuit for primality test. 
A similar  efficient construction can be done in terms of a {\sl Twin Prime} state, that 
provides the grounds for experimental counting of twin primes. The 
{\sl Prime} state can also be used to verify Goldbach conjecture. 
The entanglement properties of the {\sl 
Prime} state are directly related to counting functions of subseries of prime numbers, such as twin primes. 
Furthermore, the combination of a quantum circuit 
for primality test and the Quantum Fourier Transform allows for a counting of
prime numbers within an error which is smaller to the fluctuations allowed by the
Riemann Hypothesis.

The entanglement properties of the {\sl Prime } state remain to be explored in more detail. 
It is likely that the quantum correlations emerging from the {\sl Prime} state are profoundly related to theorems in Number Theory.

\vskip 2cm

\indent \tn{Acknowledgements.} 
The authors are grateful to J. I. Cirac, A. C\'ordoba and S. Iblisdir for helpful comments.
J. I. L. acknowledges  financial support  from FIS2011-16185, Grup de Recerca Consolidat 
ICREA-ACAD\`EMIA, and National Research Foundation \& Ministry of Education, Singapore; and G. S. 
from the grants FIS2012-33642, QUITEMAD and the Severo-Ochoa Program.

\end{document}